\documentclass[aps, prd, letterpaper, twocolumn, amsmath, amssymb, superscriptaddress, floatfix, nofootinbib, longbibliography]{revtex4-1}
\usepackage{aas_macros}
\usepackage{fancyhdr}
\usepackage[us,12hr]{datetime}
\usepackage{graphicx}
\usepackage{amssymb}
\usepackage{comment}
\usepackage{mathrsfs}
\usepackage{marvosym}
\usepackage{booktabs}
\usepackage[english]{babel}
\usepackage{amsmath,amssymb,amsbsy,amstext, amsthm, simplewick}
\usepackage{graphicx}
\usepackage[english]{babel}
\usepackage[utf8]{inputenc}
\usepackage{graphicx}
\usepackage{color}
\usepackage{xcolor}
\usepackage{amsfonts}
\usepackage{subfloat}
\usepackage{braket}
\usepackage{xcolor}
\usepackage[colorlinks=true
,urlcolor=mn
,anchorcolor=black
,citecolor=magenta
,filecolor=navyblue
,linkcolor=rindou2
,menucolor=navyblue
,pagecolor=navyblue
,linktocpage=true
,pdfproducer=medialab]{hyperref}
\usepackage{slashed}

\usepackage{aas_macros,empheq,fancybox,graphicx,multirow}

\definecolor{pc1}{rgb}{0.69, 0.25, 0.21}

\DeclareMathOperator{\tTr}{tTr}

\definecolor{rindou2}{rgb}{0.078,0.1215,0.6392}
\definecolor{mn}{rgb}{0.15, 0.35, 0.95}

\begin{document}
\title{\boldmath Tensor renormalization group study of the 3d $O(2)$ model}

\author{Jacques Bloch}
\email{jacques.bloch@ur.de}
\affiliation{Institute for Theoretical Physics, University of Regensburg, 93040
  Regensburg, Germany}
\author{Raghav G. Jha}
\email{raghav.govind.jha@gmail.com}
\affiliation{Perimeter Institute for Theoretical Physics, Waterloo, Ontario N2L 2Y5, Canada}
\author{Robert Lohmayer}
\email{robert.lohmayer@ur.de}
\affiliation{Institute for Theoretical Physics, University of Regensburg, 93040
  Regensburg, Germany}
\affiliation{RCI Regensburg Center for Interventional Immunology, 93053 Regensburg, Germany}
\author{Maximilian Meister}
\email{maximilian.meister@ur.de}
\affiliation{Institute for Theoretical Physics, University of Regensburg, 93040
  Regensburg, Germany}


\begin{abstract}
We calculate thermodynamic potentials and their derivatives for the three-dimensional $O(2)$ model
using tensor-network methods to investigate the well-known second-order phase transition.
We also consider the model at non-zero chemical potential to study the Silver Blaze phenomenon, 
which is related to the particle number density at zero temperature. Furthermore, the temperature 
dependence of the number density is explored using asymmetric lattices.
Our results for both zero and non-zero magnetic field, temperature, and chemical potential 
are consistent with those obtained using other methods.   
\end{abstract}

\maketitle

\section{Introduction}

Our understanding of quantum many-body systems has considerably improved in the past two decades 
mainly due to the refined understanding of the entangled ground state structure of systems with local 
Hamiltonians. 
Successful methods using these entanglement properties are based on the idea of tensor-network states such as matrix product states (MPS) \cite{Vidal_2003,Verstraete:2004cf,Verstraete_2006,Weichselbaum_2009}. 
They provide an efficient description of the ground states of local, gapped Hamiltonians which exhibit an 
area-law behavior. MPS 
have been applied to a wide range of problems
in different fields. These ideas have also 
been extended to two spatial dimensions 
(i.e., 2+1-dimensional quantum systems) using the generalization of MPS known as projected 
entangled pair states (PEPS), but the success has been limited. 

In addition to these methods for the continuous-time approach, an alternate 
method based on the idea of the tensor renormalization group 
(TRG) in discretized Euclidean space has also been very successful. 
This started with the pioneering work of Levin and 
Nave in two dimensions \cite{Levin:2007}. 

Both approaches have resulted in a better 
understanding of spin systems and some simple gauge theories \cite{Banuls:2019rao, 
Unmuth-Yockey:2018xak, Bazavov:2019qih, Klco:2019evd, Franco-Rubio:2019nne} 
and have been a fruitful avenue where good progress has been made. 
Though this success is impressive, it has mostly been restricted to 
two-dimensional classical or 1+1-dimensional quantum systems. 

However, the higher-order tensor renormalization group method (HOTRG) \cite{Xie_2012}, a 
Euclidean-space coarse-graining tensor method based on the higher-order 
singular value decomposition (HOSVD) \cite{DeLathauwer2000}, 
is also applicable to higher-dimensional models.
It was successfully employed to determine 
the critical temperature of the three-dimensional Ising model on a cubic lattice. 
Recently this method was used to investigate the critical behavior of the four-dimensional Ising model 
\cite{Akiyama:2019xzy}. The HOTRG method was also applied to study spin models with 
larger discrete symmetry groups such as the $q$-state Potts 
models and those with continuous global symmetries, like the classical $O(2)$ model in two 
dimensions \cite{Yu:2013sbi}, the 1+1-dimensional $O(2)$ model with chemical potential 
\cite{Zou:2014rha,Yang_2016}, and even gauge theories \cite{Bazavov:2015kka,Kuramashi:2018mmi,Kuramashi2019}.
For a review of the tensor approach to spin systems and field theory, we refer the reader to \cite{Meurice:2020pxc}.

A major drawback of the HOTRG approach is that it is very expensive in dimensions $ d \ge 3$ as the computational cost naively 
scales as $\mathcal{O}(D^{4d-1})$ with memory complexity of $\mathcal{O}(D^{2d})$ for a bond dimension $D$. 
In order to overcome this problem, new higher-dimensional tensor coarse-graining schemes, 
like the anisotropic TRG (ATRG) \cite{Adachi:2019paf} and the triad TRG (TTRG) \cite{Kadoh:2019kqk}, were recently developed. 
For ATRG the computational and storage complexity is $\mathcal{O}(D^{2d+1})$ and $\mathcal{O}(D^{d+1})$, respectively. In this work
we will use the triad method, for which the computational cost scales like 
$\mathcal{O}(D^{d+3})$ and the memory consumption like $\mathcal{O}(D^{d+2})$.%
\footnote{The complexities correspond to the original triad proposal, which uses randomized SVD (RSVD). In our implementation, we used regular SVD which makes the time complexity somewhat worse but we noted that our results are still consistent with $\mathcal{O}(D^{6})$, within errors.} Note that the improved scaling behavior comes at the cost of making additional approximations, which has to be compensated for by using larger values of $D$. 

The basic idea of the triad method is to factorize the initial and subsequent coarse-grained local tensors, which are of order $2d$ in HOTRG, into smaller tensors of order three, referred to as ``triads", by applying additional singular value decompositions (SVDs). 
For example, in $d$ dimensions, the initial fundamental tensor of order $2d$ would
decompose into ($2d-2$) triads. 
Using this factorization, all manipulations in the coarse-graining procedure can be performed at a much lower cost.

However, in this work we observed that the standard three-dimensional HOTRG algorithm can be competitive, when enhanced with some modifications. This will be illustrated in the computation of the specific heat, and also in the non-zero temperature studies with chemical potential performed using asymmetric lattices. Although the bond dimension is restricted due to the high computational cost, the improved computations of observables and the modified coarse-graining procedure for anisotropic tensors can still lead to valuable results.

Due to the ferromagnetic interaction, the $O(2)$ model in three dimensions has a continuous phase 
transition that separates a large-coupling phase with non-zero magnetization from a disordered
phase with vanishing magnetization.
This phase transition has been interpreted as
condensation of spin waves and as unbinding of vortices. 
For $\beta > \beta_{c}$, linear vortices are suppressed while they are favored for 
$\beta < \beta_{c} $. This is similar to the behavior seen in the two-dimensional version 
of the model. However, there is a crucial difference between the phase transition 
which occurs in the two-dimensional model and the one in three dimensions. 
The former is the well-known Berezinskii-Kosterlitz-Thouless (BKT) phase transition of 
infinite order, where all the derivatives of the free energy are continuous. 
However, in three dimensions, the transition is of second order, and the critical coupling can be located 
by looking at the derivatives of the free energy, which is a natural observable in 
any tensor-network calculation. 

The three-dimensional $O(2)$ model has been extensively studied using bootstrap methods and Monte Carlo (MC) methods, and critical exponents have been determined directly in the conformal field theory (CFT) limit of the model. This three-dimensional model is of special importance for many physical purposes. The $\lambda$-transition in superfluid Helium is supposed to belong to the same universality class 
as this $O(2)$ model. There is a well-known tension between theoretical/numerical predictions for the critical exponent $\alpha$ 
and the experimental values. This can be understood as follows: 
In the CFT limit, the scaling dimension $\Delta_{s}$ of a charge-zero scalar
was determined to be 1.51136(22) using bootstrap methods \cite{Chester:2019ifh}. 
From this one can compute $\nu = 1/(3-\Delta_s) = 0.67175(10)$ and the critical exponent $\alpha = 2- d\nu = -0.01526(30)$. 
These results are consistent with a recent MC study \cite{Hasenbusch:2019jkj}
which computed $\Delta_{s} = 1.51122(15)$.
On the other hand, the most precise experimental result obtained in Earth's orbit
aboard Space Transportation System (STS)-52 determined $\alpha = -0.0127(3)$ \cite{Lipa_2003}
corresponding to $\nu = 0.6709(1)$ and is in tension with the numerical estimates.
The critical coupling for the cubic 
lattice $O(2)$ model has been determined using several methods over the past three decades and we refer the reader to Table 2 of \cite{Xu:2019mvy} for a complete list. 
For example, two recent works computed 
$\beta_{\rm{c}} =  0.45416474(10)$ \cite{Hasenbusch:2019jkj} 
and $\beta_{\rm{c}} =  0.45416466(10)$ \cite{Xu:2019mvy}, respectively.
Almost all of these numerical results have been obtained 
using MC methods. 
Since tensor-network methods have been successfully used to study the $O(2)$ model in two dimensions
\cite{Yu:2013sbi, Vanderstraeten:2019frg, Jha:2020oik}, it is natural to apply these new tools also to the three-dimensional case.
Our motivation here is to carry out the first tensor 
study of the 3d $O(2)$ model
(in fact, to the best of our knowledge, the first tensor study of any three-dimensional spin model with 
continuous symmetry).

The outline of the paper is as follows: In Section \ref{sec:2}, 
we present the tensor formulation of the model using an expansion in dual variables. 
In Section \ref{sec:3}, we present our results for the pure $O(2)$ model 
both with and without an external magnetic field. Furthermore, we consider a non-zero chemical potential and compute the number density at zero and non-zero temperature and discuss the Silver Blaze phenomenon. 
We conclude the paper with a brief summary and discussion.

\section{\label{sec:2}Tensor-network formulation}
We start by considering the Euclidean action of the $O(2)$ model in the presence of an external field and chemical potential in three dimensions,
\begin{align}
\label{eq:act_1}
S = - \beta \sum_{j=1}^V \sum_{\nu=0}^2 \cos(\theta_j-\theta_{j+\hat\nu} - i \mu\delta_{\nu, 0})
- \beta h \sum_{j=1}^V \cos \theta_j ,  
\end{align}
where $j$ is a linear index defined on the cubic $N_{x}\times N_{y}\times N_{t}$ lattice with volume $V = N_{x}N_{y}N_{t}$, $\hat\nu$ denotes a unit step in direction $\nu$, $\beta$ is the coupling, $h$ is 
the external magnetic field, and the chemical potential $\mu$ only couples in the temporal direction. 
The partition function
\begin{align}
Z = \int d\Theta \,  e^{-S}
\end{align} 
is obtained by integrating over all spins $\Theta=(\theta_1,\dots,\theta_V)$ with
\begin{align}
\int d\Theta \, f(\Theta) = \int_0^{2\pi}\frac{d\theta_1}{2\pi} \dots \int_0^{2\pi} \frac{d\theta_V}{2\pi} f(\theta_1,\dots,\theta_V) . 
\end{align}
Explicitly, the partition function is
\begin{align}
Z &= 
\int d\Theta \prod_j e^{\beta h \cos\theta_j} \prod_{\nu=0}^2\, e^{\beta \cos(\theta_j-\theta_{j+\hat\nu} - i \mu\delta_{\nu, 0})} .
\end{align}
We now proceed to the dualization which results in a discrete formulation required for the
tensor-network representation. This is done using the Jacobi-Anger expansion
\begin{align}
e^{\beta\cos\theta} = \sum_{n=-\infty}^\infty I_{n}(\beta)e^{i n \theta} \,,
\end{align}
where $I_{n}(\beta)$ are the modified Bessel functions of the first kind. 
After expanding each of the exponential factors, one can integrate out all the 
spin degrees of freedom $\Theta$ to obtain an expression
for the partition function in terms of dual variables defined on the links of the lattice. 
The partition function can then be written as a complete contraction or tensor trace 
(symbolically written as  $\tTr$) of a tensor network,
\begin{align}
Z = \tTr \Big( \prod_{j=1}^V T_{lrudfb}^{(j)} \Big) \equiv \tTr (T^V),  
\label{Ztensor}
\end{align}
where the local tensor $T^{(j)}_{lrudfb}$ is the same on all lattice sites and its indices are the dual variables. In the contraction, two adjacent tensors $T^{(j)}$ and $T^{(j+\hat\nu)}$ share exactly one index, corresponding to the dual variable on their connecting link. 
For the three-dimensional $O(2)$ model the initial local tensor is
\begin{align}
\label{eq:TO2}
T_{lrudfb} &= \sqrt{I_{l}(\beta)I_{r}(\beta)I_{u}(\beta)I_{d}(\beta)I_{f}(\beta)I_{b}(\beta)e^{(u+d)\mu}} 
  \notag\\
&\times I_{l+u+f-r-d-b}(\beta h).
\end{align}
In principle, each index runs from $-\infty$ to $\infty$, but for numerical purposes, we 
truncate their ranges to size $D$ at the start and keep this size fixed during the coarse-graining procedure.\footnote{For the initial tensor, the range of each index is chosen to include the $D$ largest weights in each direction (for $h=0$). 
For $\mu = 0$, each index range is symmetric around zero, as the weights $I_{n}(\beta)$ 
have their maximum for $n=0$ and
fall off symmetrically for positive and negative $n$. For $\mu \neq 0$, the 
maximum of the temporal weights $I_{n}(\beta) e^{n\mu}$ is shifted, 
and the index range is adapted such that it still covers the $D$ largest weights.}

For $h=0$, the tensor enforces a Kronecker delta on the backward and forward indices,
\begin{align}
\lim_{h\to 0} I_{l+u+f-r-d-b}(\beta h) = \delta_{l+u+f}^{r+d+b}\,,
\label{h=0_limit}
\end{align}
corresponding to the global $O(2)$ symmetry of the action, which ensures that the total directed flux that enters any site vanishes.

\begin{figure}
\centering 
\includegraphics[width=0.48\textwidth]{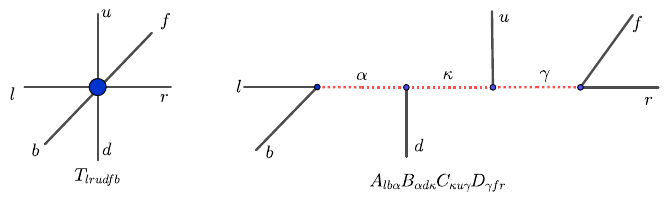}
\caption{\label{fig:1}The representation of the initial tensor and its decomposition into triads for a three-dimensional system. 
The contracted indices are shown by red dashed lines.}
\end{figure}

The basic principle of the HOTRG algorithm is to perform successive coarse-graining 
operations in order to evaluate the partition function \eqref{Ztensor}. Each of these coarse-graining contractions squares the dimension of the indices perpendicular to the contraction direction, as indices with dimension $D$ from two adjacent tensors are combined into a ``fat'' index of dimension $D^2$. To avoid a blow up of the dimension of the coarse-grained local tensor, the algorithm then applies an HOSVD approximation \cite{DeLathauwer2000,Xie_2012} after each coarse-graining step, which truncates the dimension of each fat index back from $D^2$ to $D$. Consequently, the local tensor remains of dimension $D^{2d}$, and coarse graining is performed until only one single tensor remains. Finally, this remaining tensor is contracted over its corresponding backward and forward indices to yield the partition function $Z$. Thermodynamic observables are computed either by taking finite-difference numerical derivatives of the partition function \eqref{Ztensor} or by using an impurity method, where the derivative of $\ln Z$ is directly applied to the tensor network, see \eqref{Mwithimp}.
Our computations were performed using both the triad method and the standard HOTRG method. The approximate decomposition of the local tensor in triads, which is used both for the initial and coarse grained tensor, is shown in Fig.\ \ref{fig:1}.

\section{\label{sec:3}Results} 
\subsection{\label{subsec:3A}\boldmath $\mu = 0$, $h=0$} 

\begin{figure}
\centering 
\includegraphics[width=0.48\textwidth]{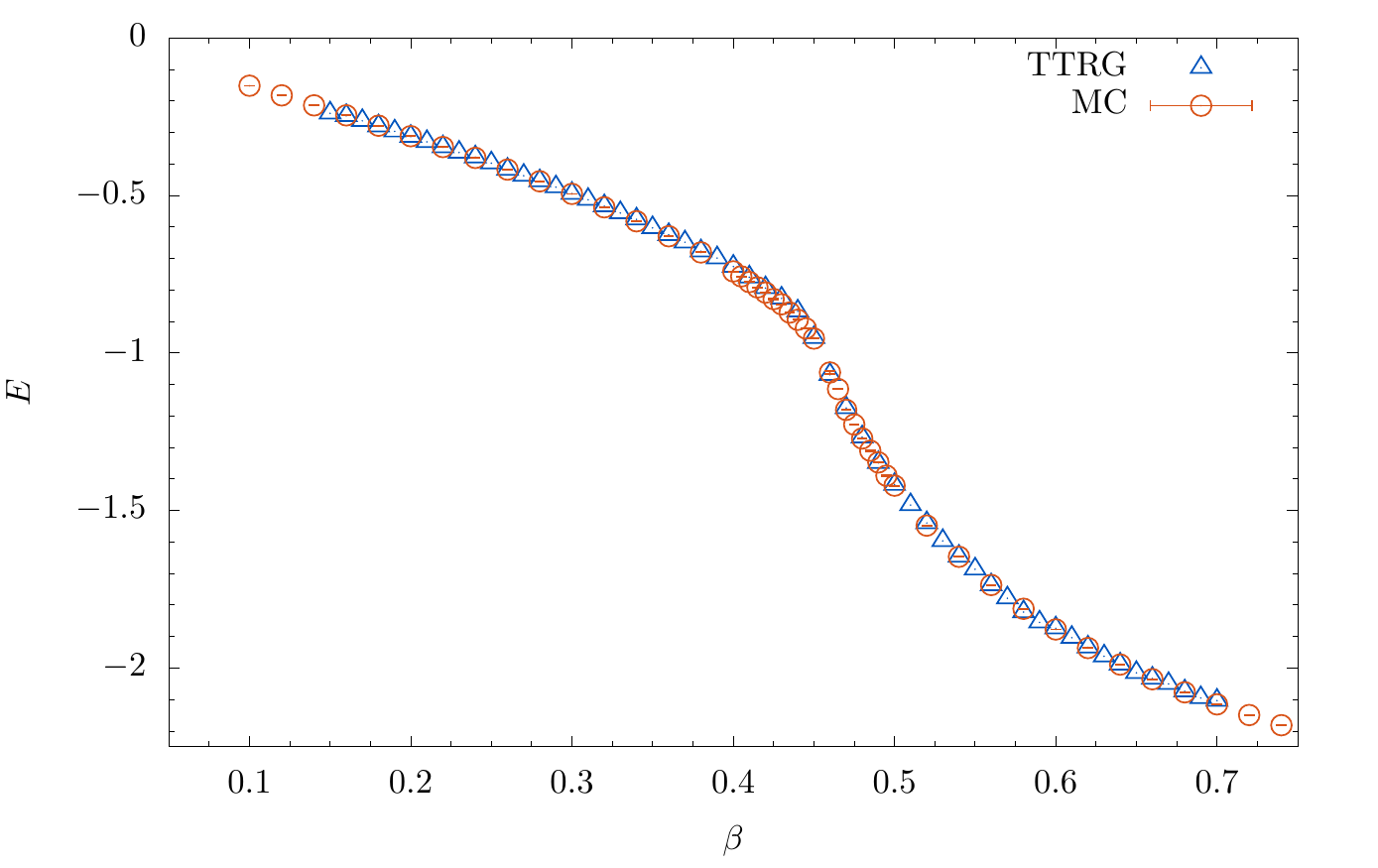}
\caption{\label{fig:2}The internal energy 
obtained using triad TRG with $D=50$, and finite differences with $\Delta\beta=0.02$, agrees with 
the MC results on a lattice volume of $32^{3}$.}
\end{figure}

In this subsection, we will discuss the $O(2)$ model
without chemical potential or magnetic field. 
For $\mu=0$ and $h=0$ the initial local tensor (\ref{eq:TO2}) simplifies to
\begin{align}
\label{eq:T-mu=h=0}
T_{lrudfb} &= \sqrt{I_{l}(\beta)I_{r}(\beta)I_{u}(\beta)I_{d}(\beta)I_{f}(\beta)I_{b}(\beta)} \,
\delta_{l+u+f}^{r+d+b}.
\end{align} 
In this case, all the external triad legs carry the same weights. One of the observables we compute is the ``internal energy''\footnote{A name we use for the average action density (up to a factor of $\beta$) in analogy to classical statistical systems.}
\begin{equation}
	E = -\frac{1}{V} \frac{\partial \ln Z}{\partial \beta}.
\end{equation}
We show that the results obtained using the triad tensor method agree with those from the MC approach, as illustrated in Fig.\ \ref{fig:2}.

\begin{figure}
\centering 
\includegraphics[width=0.48\textwidth]{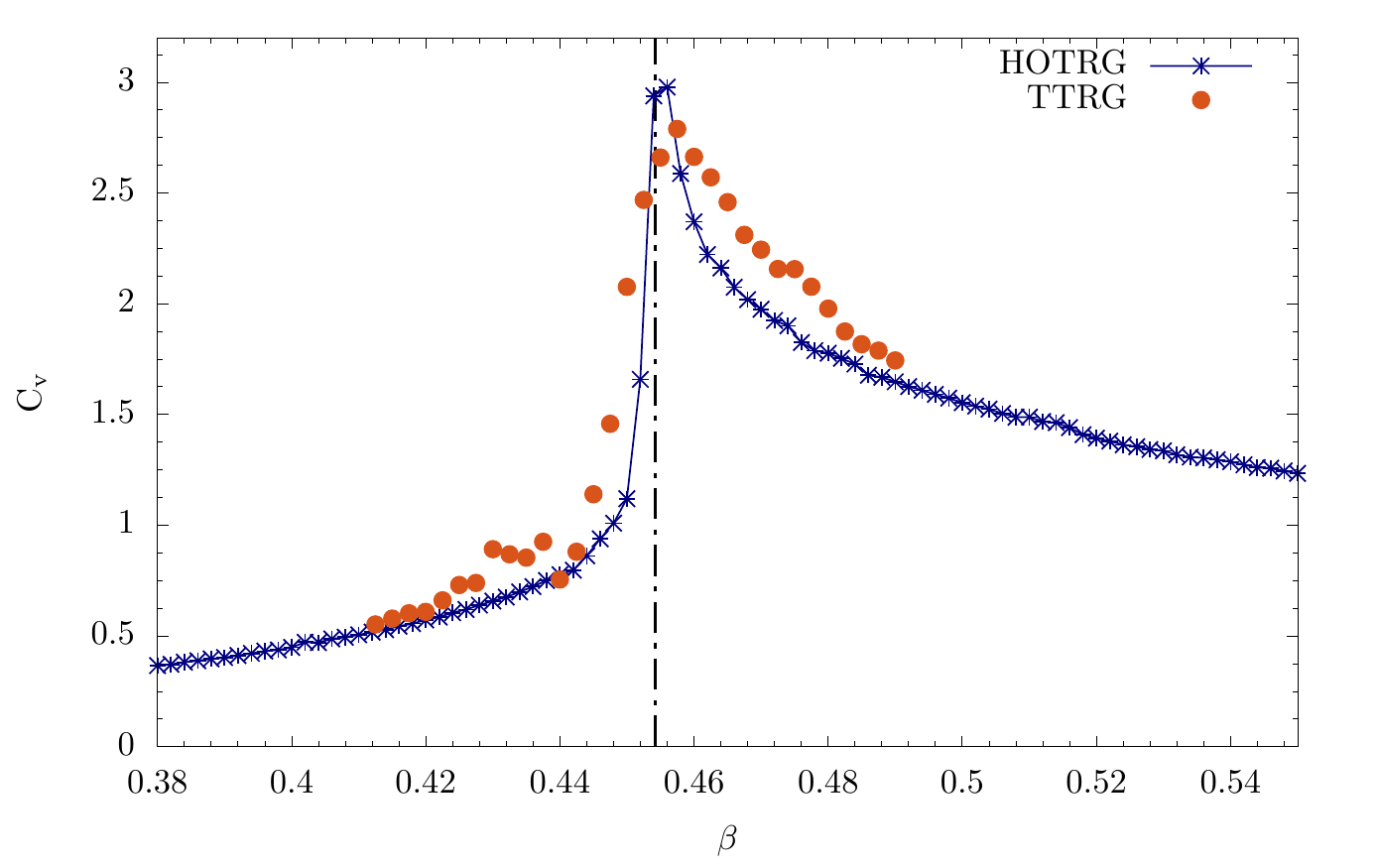}
\caption{\label{fig:3}
The specific heat capacity as a function of $\beta$ for a $32^{3}$ lattice volume. 
The triad data (orange) are computed with $D=72$ using a second-order finite difference of $\ln Z$ with 
step size $\Delta\beta=0.01$. The HOTRG data (blue) used $D=15$ and are computed with a 
\textit{stabilized} second-order finite-difference scheme with $\Delta\beta=10^{-6}$. 
The peak of $C_{\rm v}$ suggests that the critical coupling is between $\beta = 0.45$ and $\beta=0.46$.
For reference, we show the infinite-volume MC result $\beta_{\rm{c}} =  0.454165$ from \cite{Xu:2019mvy} by the black dashed line.}
\end{figure}

In order to determine the critical coupling, we compute the second $\beta$ derivative of the logarithm of the partition function to determine
the ``specific heat",
\begin{equation}
	C_{\rm v} = \frac{\beta^2}{V} \frac{\partial^2 \ln Z}{\partial \beta^2}.  
\end{equation}
Our results shown in Fig.\ \ref{fig:3} clearly indicate that there is a peak in the specific heat corresponding to the second-order phase transition in this model. The location of the peak is consistent 
with high-precision results of earlier studies. 
The triad data are computed with $D=72$ using second-order finite differences with step size $\Delta\beta=0.01$.
Decreasing the step size to reduce discretization errors is problematic as the systematic errors on $\ln Z$ cause large fluctuations on the standard finite-difference derivatives, and one would 
require a larger bond dimension ($D$) or more sophisticated numerical derivative computations to achieve a 
precise determination of the peak.
The results of such an improvement for the original HOTRG method can be seen in Fig.\ \ref{fig:3} for $D=15$, where we get a smooth behavior for the specific heat, including the steep phase-transition region. These data were obtained using a \textit{stabilized} second-order finite-difference scheme with step size reduced to $\Delta\beta=10^{-6}$.
The stabilized finite-difference scheme was developed to avoid jumps between values of $\ln Z$ computed on close-by parameter values required for the evaluation of finite differences. Typically such jumps are caused by degenerate singular values or level crossings of singular values, leading to discontinuous changes of the vector subspaces used to truncate the coarse-graining tensors. The stabilization uses a heuristic approach that operates on the singular vectors of HOTRG to maximize the overlap between the adjacent vector subspaces (adjacent under a small change of $\beta$ in this case). These stabilized subspaces then improve the smoothness of $\ln Z$ for adjacent parameter values used to compute finite-difference derivatives. The application of stabilized finite differences to triads is more subtle and left for future work.
Note that observables can also be computed using the impurity method (e.g., first order for the energy, second order for the specific heat). Although this method yields smoother data (which does not necessarily mean more accurate) than the finite difference method, it has an additional systematic error because the same matrices of singular vectors are used to truncate the pure and impure tensors.

\subsection{\boldmath $\mu = 0$, $h \neq 0$} 
\begin{figure}
\centering 
\includegraphics[width=0.48\textwidth]{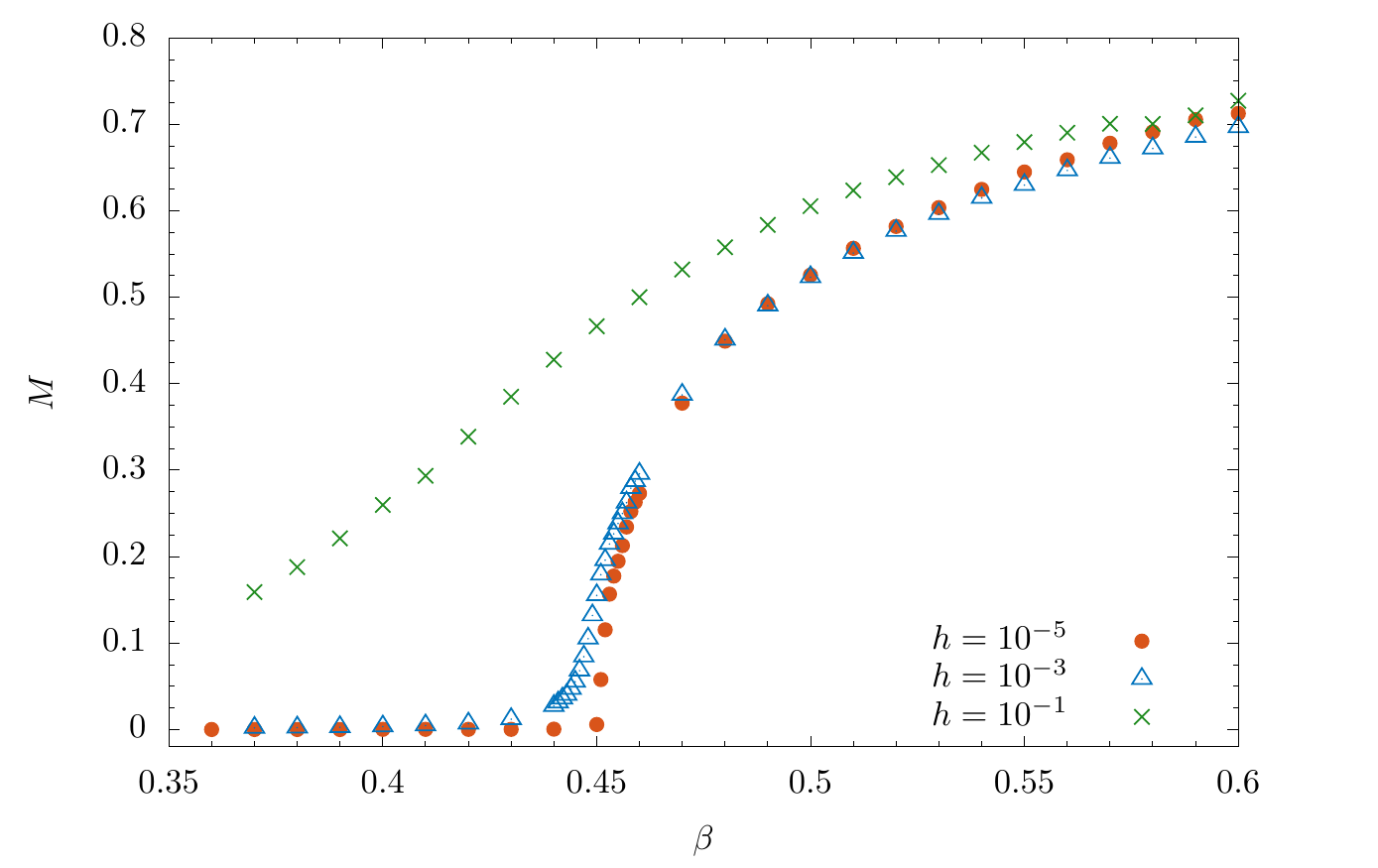}
\caption{\label{fig:4}The magnetization of 
the $O(2)$ model for different external magnetic fields. We see that for 
sufficiently small symmetry-breaking field, the magnetization rises sharply 
around $\beta_{c}$. These results are obtained on a lattice of volume  
$(2^{13})^{3}$ with $D=30$.
}
\end{figure}

In this subsection, we study the model in the presence of a small symmetry-breaking external field $h$. 
The global $O(2)$ symmetry is broken and the partition function is given by
\begin{align}
	Z &= 
	\int d\Theta \prod_{i} e^{\beta h \cos\theta_i} \prod_{\nu = 0}^{2} e^{\beta \cos(\theta_i-\theta_{i+\hat\nu})} .
\end{align}
One can compute the magnetization by either taking a numerical derivative of $\ln Z$ with respect to
$h$ or by inserting an impurity tensor in the tensor network.
Here we use the latter method with the impurity tensor given by
\begin{align}
\label{eq:TO2withH}
\widetilde{T}_{lrudfb} &= \frac{1}{2} \sqrt{I_{l}(\beta)I_{r}(\beta)I_{u}(\beta)I_{d}(\beta)I_{f}(\beta)I_{b}(\beta)}   \notag\\
& \hspace{-2mm}\times \Big( I_{l+u+f-r-d-b+1}(\beta h) + I_{l+u+f-r-d-b-1}(\beta h) \Big) 
.
\end{align}
From $\widetilde{T}_{lrudfb}$ and 
$T_{lrudfb}$ we can then compute the magnetization density as
\begin{equation}
M = \frac{1}{V} \sum_{i} \langle \cos\theta_i \rangle = \frac{1}{\beta V} \frac{\partial \ln Z}{\partial h} = \frac{\tTr(\widetilde{T}T^{V-1})}{\tTr(T^V)}. 
\label{Mwithimp}
\end{equation}
The results we obtained for the average magnetization density are shown in Fig.\ \ref{fig:4}. For $\beta < \beta_{c}$, the spins are randomly distributed 
and average to zero, while for $\beta > \beta_{c}$ 
they prefer to align, resulting 
in a non-zero net magnetization. As we explore smaller $h$, we see that the change of behavior is consistent with the 
critical coupling obtained from the peak of the specific heat.

\subsection{\boldmath $\mu \neq 0$, $h=0$} 

In this subsection, we consider the $O(2)$ model in the presence of a chemical potential, for which the action was already given in \eqref{eq:act_1}. This generalization is a problem for standard MC methods because 
the probability distribution in the partition function becomes complex and the sign problem is encountered
(like for QCD at non-zero baryon density). Some numerical methods devised to circumvent the sign problem are 
reweighting, complex Langevin, thimbles, density of states, and dual variables. Reweighting enables the 
use of importance sampling MC, but only at an exponential cost, which makes it unusable for any practical purpose. 
The complex Langevin method uses a complexification of the spin degrees of freedom, 
however, measurements on the enhanced partition function are only equivalent to those on the original one 
if specific conditions concerning the probability distribution of the drift term in the complex plane are met 
\cite{Aarts:2011ax,Nagata:2016vkn}. For the three-dimensional $O(2)$ model, the method does 
not satisfy these conditions in the disordered phase ($\beta \leq \beta_c$) 
and the method produces erroneous results \cite{Aarts:2010aq}.

The method of choice to tackle the sign problem in the three-dimensional $O(2)$ model  
is to introduce dual variables, as discussed in Sec.\ \ref{sec:2}, and integrate out the original spin degrees of freedom. 
The ensuing partition function is free of a sign problem, even in the presence of a chemical potential. 
Once rewritten in this way the partition function can be simulated by the worm algorithm \cite{Prokof_ev_2001}, 
as was done successfully in \cite{Banerjee:2010kc,Langfeld:2013kno}. 

Once reformulated in terms of dual variables, it turns out that the partition function can also 
be interpreted as a tensor network \eqref{Ztensor}, and tensor-network methods can be applied in a 
straightforward way, as discussed in Sec.\ \ref{sec:2}.
The only effect of the chemical potential is to modify the tensor entries depending on the value of their temporal indices, 
\begin{align}
\label{eq:T-mu}
T_{lrudfb} &= \sqrt{I_{l}(\beta)I_{r}(\beta)I_{u}(\beta)I_{d}(\beta)I_{f}(\beta)I_{b}(\beta)e^{(u+d)\mu}} \notag\\
&\times\delta_{l+u+f}^{r+d+b}.
\end{align} 
Note that even if a sign problem would remain after dualization, which would require reweighting
in the worm algorithm, this would not be an issue for the tensor-network method which is 
deterministic in its construction and remains unaffected by such inconveniences, 
at least concerning the methodology.

One of the interesting observables at non-zero chemical potential is the
particle number density (or charge density) defined as
\begin{align}
    \rho = \frac{1}{V} \frac{\partial \ln Z}{\partial \mu}.
\end{align}
In this subsection, we will investigate two important aspects of the $O(2)$ model at non-zero chemical potential: the Silver Blaze phenomenon at zero temperature and the temperature dependence of $\rho$, which is studied using asymmetric lattices.

As will be detailed below, we observe that for symmetric lattices the number density remains zero up to some threshold $ \mu = \mu_{c}$ and then becomes non-zero, confirming the results of Ref.\ \cite{Langfeld:2013kno}. This is a phenomenon 
occurring at strictly zero temperature since there the thermodynamic quantities are independent 
of $\mu$ when $ \mu < \mu_{c}$, i.e., as long as $\mu$ is below the mass of the lightest 
excitation (or mass gap). In this case, no particle excitations can be generated and the particle 
number density is independent of $\mu$. This has been dubbed as the 
\emph{Silver Blaze phenomenon}\footnote
{The name is inspired from ``The Adventure of Silver Blaze", one of Sherlock Holmes short stories written 
by Sir Arthur Conan Doyle and first published in December 1892.
In this story, Holmes used the ``curious incident" of a dog doing nothing in the night time as a key clue to solve
the mystery of a missing horse named ``Silver Blaze" and the death of its trainer. In this context, the issue is to understand 
the $\mu$-independence of physical quantities, i.e., why the chemical potential 
does nothing for $\mu < \mu_{c}$ even when it is in the action.} 
in studies of various lattice theories
\cite{Cohen:2003kd}.  

The Silver Blaze phenomenon is especially hard to reproduce numerically as it is closely related to 
the cancellations in the original partition function which also lead to the sign problem. 
This is seen in MC simulations when reweighting from the phase quenched to the full theory. 
In the phase quenched theory the complex action is replaced by its real part, i.e., the weights in the original partition function are replaced by their magnitude.
The phase quenched theory has no Silver Blaze, i.e., the particle number steadily increases with $\mu$. 
In this case, the Silver Blaze property of the full theory should emerge 
from large cancellations of the phase, however, only at an exponential cost \cite{Aarts:2013bla}. 
Such reweighting simulations of the $O(2)$ model clearly show that the Silver Blaze is beyond reach 
using such methods.

\begin{figure}
\centering 
\includegraphics[width=0.48\textwidth]{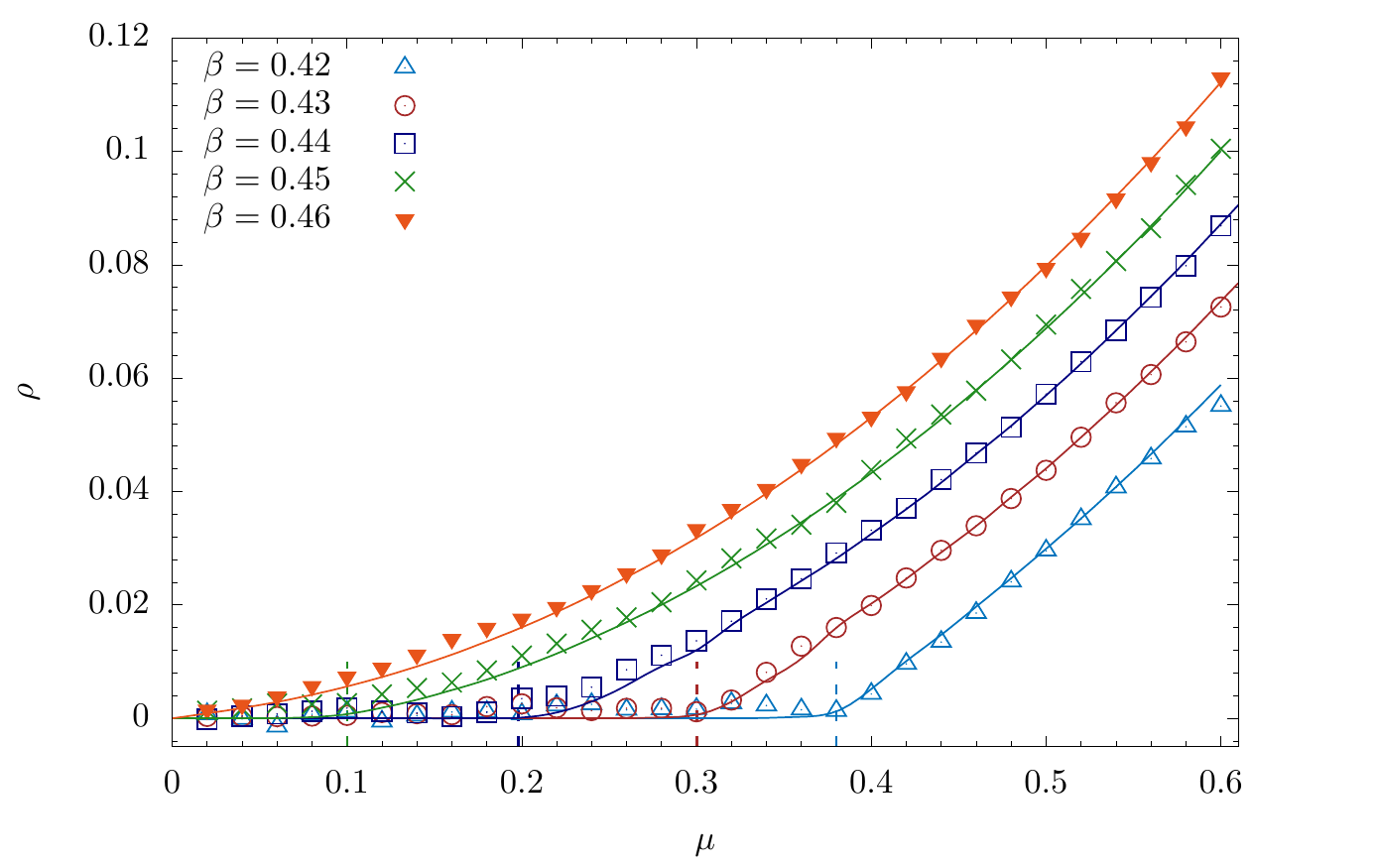}
\caption{\label{fig:5}We compare the results 
	obtained using triad TRG (symbols) with $D=50$ and worm algorithm (smooth lines) 
	for the dependence of $\rho$ on $\mu$ for some values of the coupling $\beta$ on both sides (phases) 
	of the critical coupling on a lattice of size $64^{3}$. We mark the threshold value
$\mu_c$ which is related to the mass gap. It is clear that the mass gap 
decreases (and correlation length increases) 
as we go from $\beta=0.42$ to $\beta=0.45$
and would go to the CFT limit as $ \beta \to \beta_{c} \approx 0.45417$.}
\end{figure}

However, as can be seen in Fig.\ \ref{fig:5}, 
the Silver Blaze can be nicely reproduced, both by the worm algorithm and by the tensor method 
used in this work, and the results from both methods are in good agreement. 
In our tensor-network calculations, $\rho$ is computed using finite differences of $\ln Z$.
We used a relatively small lattice size of $64^3$, which is primarily 
due to the large cost of the worm algorithm as the volume increases. For the range of $\beta$ values considered in the figure, this does not affect the results as the correlation lengths are small compared to the box size. 
This was also verified using tensor computations with volumes up to $1024^3$ which gave results similar to $64^3$. In Fig.\ \ref{fig:5}, 
we show how the threshold $\mu_c$ varies with the coupling $\beta$ as we approach the continuum limit, i.e., 
$\beta\to\beta_c$ where the lattice spacing $a\to 0$. 
As expected, we see that in the bare theory the threshold tends to zero as $ \beta \to \beta_{c}$. 
If we were to renormalize the lattice quantum field theory (see \cite{Langfeld:2013kno}) 
and set the lattice spacing in physical units, the physical chemical potential $\mu_\text{ph}=\mu/a$, would have a 
threshold $(\mu_c)_\text{ph}$ corresponding to the particle mass, independently of the value of $\beta$ in the 
vicinity of $\beta_c$ (up to discretization errors).

We can also use the tensor methods to study the $O(2)$ model at non-zero temperature.
For this we note that the extent of the Euclidean time axis is inversely proportional to the 
physical temperature, i.e., $T=1/(N_{t} a)$. The temperature can be set by varying the number 
of temporal sites $N_t$, and can be further fine-tuned by changing the coupling $\beta$ which
determines the lattice spacing. 

In standard HOTRG, the iterative coarse-graining procedure alternates between 
the different directions, here $t,x,y$, until the complete network has been reduced to a single tensor. 
This is a natural (although not necessarily best) coarse-graining order for an isotropic tensor on a 
symmetric lattice ($N_t=N_{x,y}$).

In the case of asymmetric lattices ($N_t \neq N_{x,y}$) a different strategy is often employed to compute results for varying values of $N_t$, i.e., temperatures, in an efficient way. The procedure consists of performing all spatial contractions on a single time slice to produce a time transfer matrix \cite{Zou:2014rha}. 
This time transfer matrix is then multiplied to itself to attain the required number of time slices.
Unfortunately, it turns out that such a procedure only converges to the correct result, obtained using the worm algorithm, for large $N_t$ (zero temperature) and often yields substantial deviations for non-zero temperatures. An alternative procedure is used in Ref.\ \cite{Kuramashi:2018mmi} where finite temperature results are obtained in 2+1-dimensional $Z_2$ gauge theory for small $N_t$ by completely contracting the temporal direction first, and then coarse-graining the remaining spatial directions.

For anisotropic tensors, e.g., caused by a chemical potential, special care has to be taken to the coarse-graining order, i.e., the order in which the directions get contracted, to avoid large truncation errors. 
We therefore developed a method that implements an improved contraction order (ICO). This new method dynamically selects the next contraction direction to minimize the local truncation error. Its flexibility also makes it very
useful for the treatment of asymmetric lattices and the method performs well 
for both small and large $N_t$.\footnote{For small anisotropy (small chemical potential) and $N_t<N_{x,y}$ the ICO procedure typically alternates the coarse graining between all directions until the time direction is completely contracted. Then the tensor is reduced to an effective two-dimensional spatial tensor and the remaining spatial contractions are performed, alternating over $x$ and $y$ like in standard two-dimensional HOTRG. This specific procedure can also be ported to the triads.}
The ICO method was implemented as an enhancement of the standard HOTRG method. It was not yet implemented for the TTRG method because of the peculiar anisotropy of the triad factorization.

\begin{figure}
\includegraphics[width=0.48\textwidth]{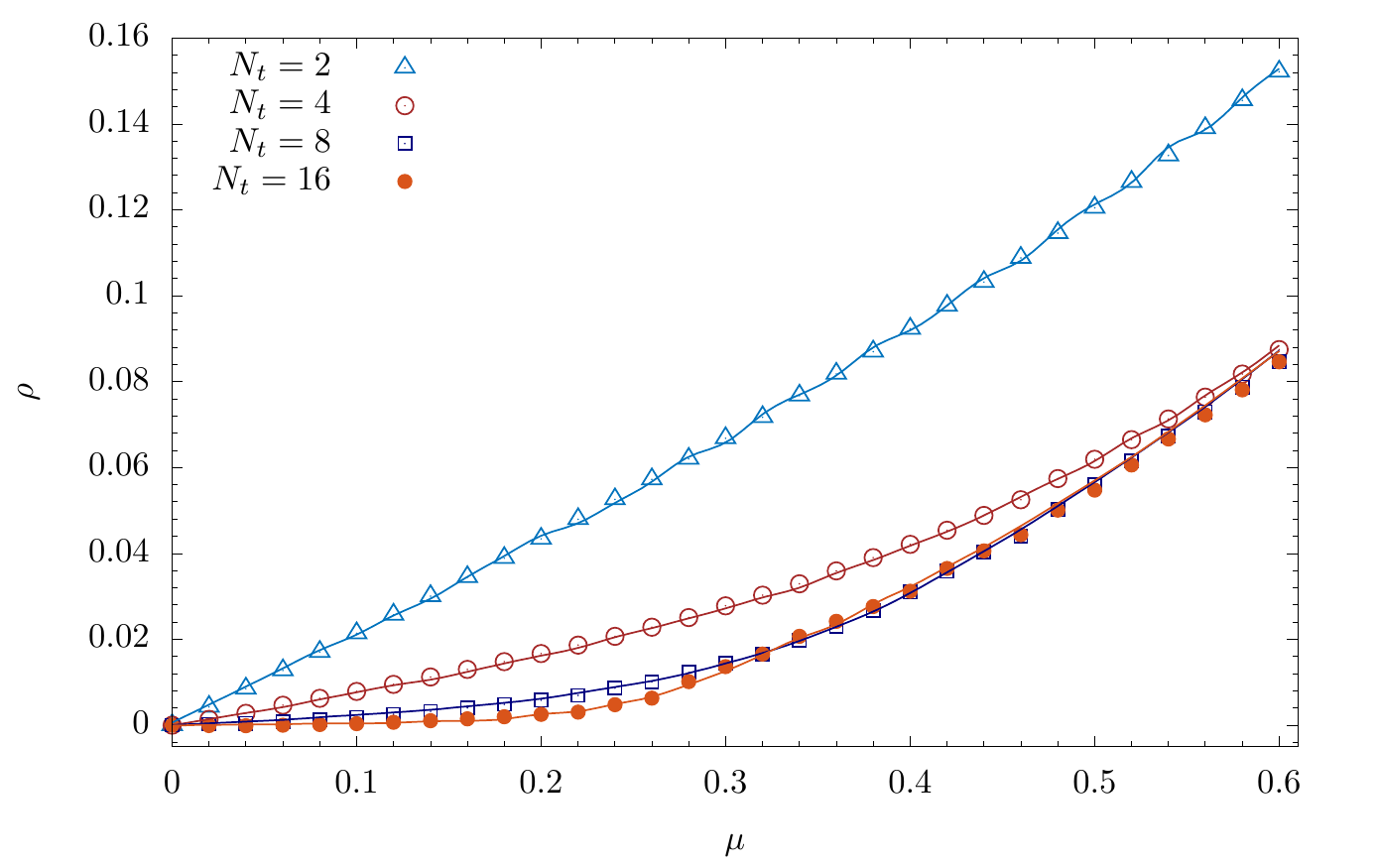}
\caption{We use HOTRG with $D=13$, improved contraction order and stabilized finite differences 
to compute the particle density $\rho$ for a $64^2\times N_t$ lattice with $N_t=2,4,8,16$ (symbols) at $\beta=0.44$ 
and compare to the results obtained using worm algorithm (smooth lines). There is clear indication that as we move towards
zero temperature, the behavior we see in Fig.\ \ref{fig:5} starts to emerge.}
\label{Fig:Nt}
\end{figure}

To validate the non-zero temperature tensor results we used the worm algorithm 
\cite{Prokof_ev_2001} at non-zero $\mu$ and find good agreement. This is illustrated in
Fig.\ \ref{Fig:Nt} where we show the temperature dependence of the 3d $O(2)$ model by studying
the system on a $64^2 \times N_t$ lattice for $N_t=2,4,8,16$. 
The tensor results were obtained using the ICO enhanced HOTRG method with $D=13$. The particle number density was computed using a 
stabilized finite-difference scheme (see Subsec.\ \ref{subsec:3A}), and 
tensor manipulations were performed using the TBLIS library \cite{matthews2016highperformance}.

\section{Summary and Discussion}

In this work, we have carried out the first tensor-network study of the 3d classical 
$O(2)$ model at both zero and non-zero magnetic field, chemical potential, and temperature. 
The results obtained for the internal energy and the specific heat are consistent with MC data. 
However, our determination of the critical coupling is several orders of magnitude less precise 
than state-of-the-art MC results. We calculated the magnetization in the presence of a small magnetic field by 
inserting an impure tensor. At non-zero chemical potential, we were able to reproduce the Silver Blaze 
phenomenon at zero temperature. We considered non-zero temperature by varying 
the temporal extent of the lattice and computed the particle density at non-zero chemical potential. 
Our results agree with those obtained with the worm algorithm.

In the appendix, we discuss the convergence of $\ln Z/V$ with the bond dimension $D$. 
We expect that this convergence will play a key role in a more precise determination of $\beta_{c}$ 
and in exploring the corresponding field-theory limit in the future. 
To this end, improved coarse-graining schemes will have to be developed.
Such improvements will also be useful to explore other interesting spin models in the future.

\section*{Acknowledgements}
We thank Judah Unmuth-Yockey and Michael Nunhofer 
for discussions. Some of the numerical computations were done on 
Symmetry which is Perimeter's HPC system. RGJ is supported by a postdoctoral 
fellowship at the Perimeter Institute for Theoretical Physics. Research at Perimeter Institute 
is supported in part by the Government of Canada through the Department of Innovation, 
Science and Economic Development Canada and by the Province of Ontario through the 
Ministry of Economic Development, Job Creation and Trade.

\section*{Appendix}
It is a known problem that the truncations used in tensor-network
methods sometimes lead to drastic modifications of the properties of the model 
whose thermodynamic behavior one intends to study.
In this appendix, we investigate the convergence of $\ln Z/V$ with the local bond dimension $D$ in the triad approximation of HOTRG,
in the large-volume limit for 
the three-dimensional cubic Ising and $O(2)$ models.
We tune the couplings close to their critical values to make the 
dependence on $D$ prominent. 
This is illustrated in Figs.\ \ref{fig:7} and \ref{fig:8}. 
The shaded areas enclose the various fits to the data (corresponding to various fit ranges and different fit formulas, including the Ansatz $a+bD^{-c}$). 
The extrapolation to $D\to\infty$ can be read off from the intercept with the vertical axis.
The convergence for the Ising model is faster than for the $O(2)$ model, which may
hint to a different efficiency of tensor methods for systems with discrete and continuous symmetries.
In order to explore the field-theory limit and for the determination of the critical 
exponents, the infinite-$D$ value of $\ln Z/V$ is required to good accuracy. 
It seems that the current tensor computations
are still somewhat far away from that desired limit. 
An advance in constructing better algorithms that converge faster 
without increasing time or memory complexity would be desirable in the future.

The numerical computations were mostly performed on a 2.4 GHz machine 
with about 180 GB of memory on a single core using the highly optimized $\texttt{opt\_einsum}$
Python module for tensor contractions \cite{Smith:2018aaa}. We explored a maximum of
$D = 82$ for the $O(2)$ model which took about 62 hours for a lattice of volume 
$(2^{15})^3$ and about 18 hours for a lattice volume $(2^{5})^3$. 
We found that the computation time for the triad method scaled as $D^6$ within errors, 
even though our implementation of the algorithm asymptotically scales as $D^7$. 

\begin{subfigures}
  \begin{figure}
    \begin{minipage}[t]{0.48\textwidth}
  \includegraphics[width=0.99\textwidth]{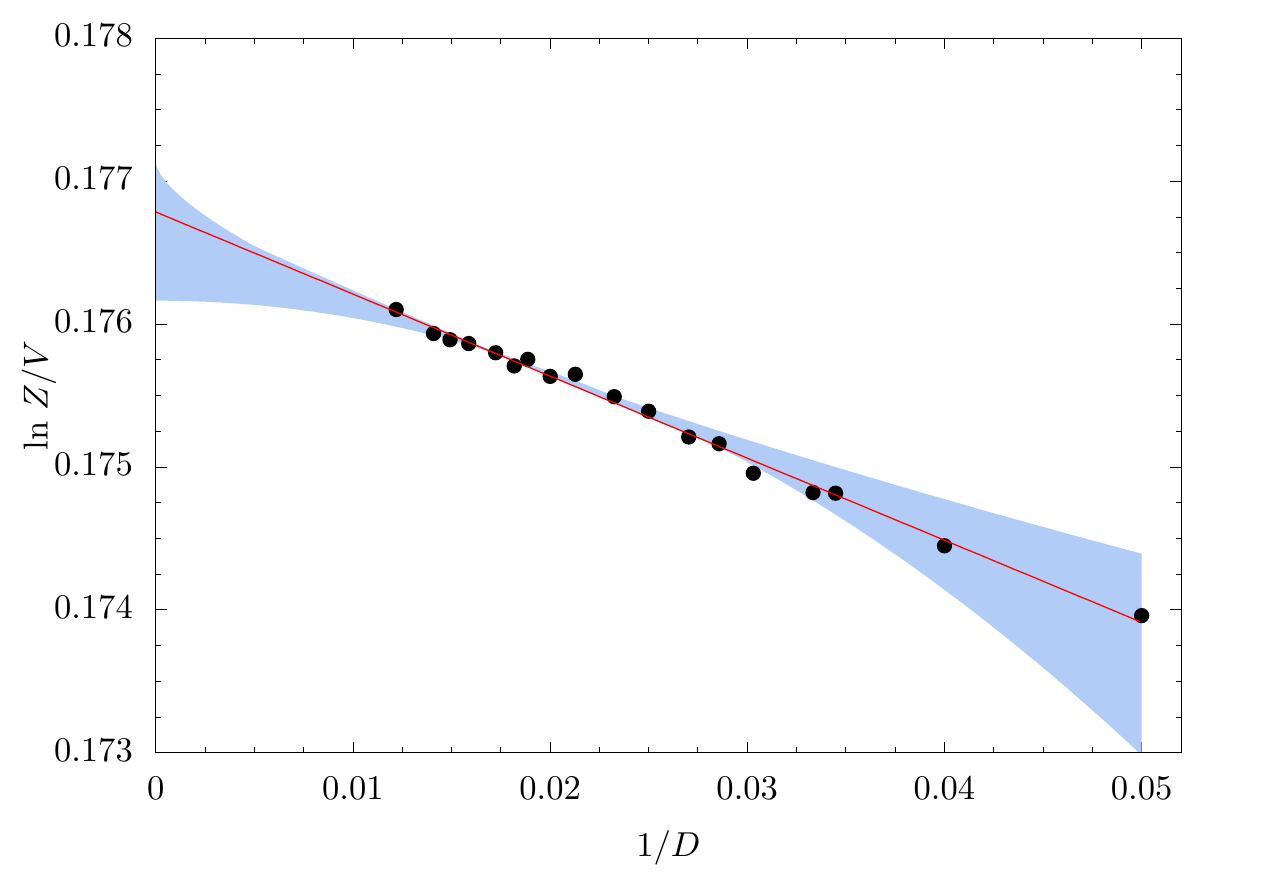}
  \caption{The dependence of $\ln Z/V$ on the local bond dimension $D$ on a lattice of volume
    $(2^{15})^{3}$ at $\beta = 0.45417$ for the 3d $O(2)$ model obtained using the triad method. The red line shows the result of a linear fit using all data points.}
  \label{fig:7}
    \end{minipage}
    \hfill
    \begin{minipage}[t]{0.48\textwidth}
  \includegraphics[width=0.99\textwidth]{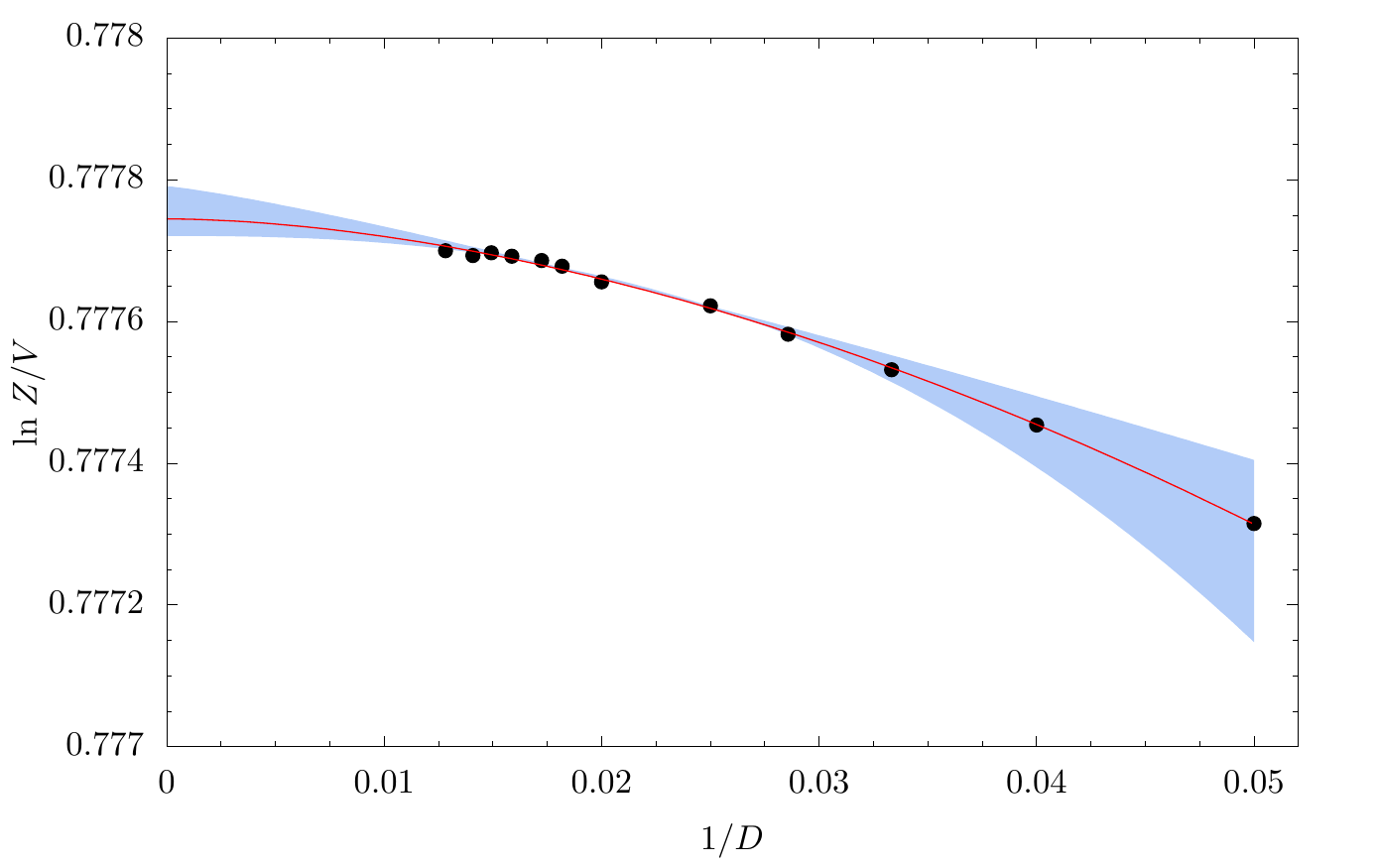}
  \caption{The dependence of $\ln Z/V$ on $D$ on a lattice of volume
  $(2^{15})^{3}$ at $T = 4.5115$ for the 3d classical Ising model using triads. The red line shows the result of a quadratic fit using all data points. The current best estimate of $T_{c}$ on a cubic lattice is 4.5115247 within errors \cite{Xu:2018hwn}.}
  \label{fig:8}
  \end{minipage}
\end{figure}
\end{subfigures}

\newpage 
\bibliographystyle{utphys}
\raggedright
\bibliography{v1}
\end{document}